\documentstyle{lamuphys}
\makeatletter
\let\chapter\hid@chapter
\makeatother
\def\ss{\sigma_0}
\def\re{R_e}
\def\lb{L_B}
\def\ie{\langle I \rangle _e}
\def\ku{k_1}
\def\kd{k_2}
\def\kt{k_3}
\def\ms{M_*}
\def\mh{M_h}
\def\mdr{{\cal R}}
\def\ml{\Upsilon_*}
\def\uss{U_{**}}
\def\wsh{W_{*h}}
\def\ck{C_K}
\def\dvm{R^{1/m}}

\def\err{\Theta}
\def\mi{M_i}
\def\mr{M_R}
\def\mto{M_{TO}}
\def\msun{M_{\odot}}
\def\min{M_{inf}}

\def\rcs{r_*}
\def\rch{r_h}
\def\rt{r_t}
\def\ra{r_a}

\begin{document}
\pagenumbering{arabic}
\title{The Physical Origin of the Fundamental Plane (of Elliptical Galaxies)}

\author{Luca\,Ciotti}
\institute{Osservatorio Astronomico di Bologna,\\
via Zamboni 33, 40126 Bologna, Italy}

\maketitle

\begin{abstract}
I review the basic problems posed by the existence of the Fundamental Plane, 
and discuss its relations with the Virial Theorem (VT). Various possibilities
are presented that can produce the observed uniform departure from 
{\it homology} (structural, dynamical) and/or from a constant {\it stellar} 
mass-to-light ratio. The role of orbital anisotropy and its relation with the 
FP thickness are also discussed. None of the explored solutions -- albeit 
formally correct -- are easily acceptable from a physical point of view, due 
to the ever-present problem of the required fine-tuning.
\end{abstract}
\section{Observational Facts}

Three are the main global observables of elliptical galaxies (Es): the central
projected velocity dispersion $\ss$, the effective radius $\re$, and the mean 
effective surface brightness within $\re$, $\ie=\lb/2\pi\re^2$. It is well 
known that Es do not populate uniformly this three dimensional parameter 
space; they are rather confined to a narrow logarithmic plane (Dressler et al.
 1987; Djorgovski and Davis 1987), thus called the {\it Fundamental Plane} 
(FP). For example, for Virgo ellipticals
\begin{equation}
\re\propto \ss^{1.4}\,\ie^{-0.85}\enspace .\label{VFP}
\end{equation}
Bender, Burstein, and Faber (1992, hereafter BBF) have introduced the $k$ 
coordinate system, in which the new variables are a linear combination of the 
observables:
\begin{equation}
\ku \equiv (2\log\ss+\log\re)/\sqrt{2}\enspace,\quad
\kd \equiv (2\log\ss+2\log\ie-\log\re)/\sqrt{6}\enspace,\label{k12}
\end{equation}
\begin{equation}
\kt \equiv (2\log\ss-\log\ie-\log\re)/\sqrt{3}\enspace.\label{k3}
\end{equation}
In the new space, the $\ku$-$\kt$ plane provides an almost edge-on view of the 
FP, and $\kt=0.15\ku+{\rm const.}$ (Fig. 1 of BBF). Here I consider some of 
the implications of the two main properties of the FP of Virgo and Coma 
cluster ellipticals: 1) the FP is remarkably thin, with a 1-$\sigma$ 
dispersion $\sigma(\kt)\simeq\pm 0.05$ (Bender, private communication); and 2)
this thickness is nearly constant along the FP. First, I will try to clarify 
some frequently misunderstood points about the relations between the FP and 
the VT, and the meaning of structural/dynamical non-homology.  

\section{The FP and its relation with the VT}

The characteristic dynamical time of Es (e.g., within $\re$) is 
\begin{equation}
T_{dyn}\simeq (G<\rho>_e)^{-1/2}\approx 10^8 {\rm yrs}\enspace, \label{tdyn}
\end{equation}
and their collisionless relaxation time is of the same order (Lynden-Bell 
1967), i.e., both are short with respect to the age of Es. As a consequence, 
only highly perturbed galaxies are presumably caught in a non-stationary 
phase. Stationarity is a {\it sufficient} condition for the validity of the VT,
and so for Es the Virial Theorem holds. For a galaxy of total stellar mass 
$\ms$ embedded in a dark matter halo of total mass $\mh$, the scalar VT can be 
written as
\begin{equation}
<V_*^2>=(G \ml\lb/\re)\times
(|\tilde\uss| + \mdr |\tilde\wsh|)\enspace, \label{SVT}
\end{equation}
where $\mdr\equiv\mh /\ms$ and $\ml=\ms/\lb$ is the {\it stellar} mass-to-light
ratio, here defined using the galaxy total {\it blue} luminosity. The 
dimensionless functions $\tilde\uss$ and $\tilde\wsh$ are the stellar 
gravitational self-energy and the interaction energy between the stars and the
dark matter halo. They depend only on the stellar density profile and on the 
relative distribution of the stellar matter with respect to the gradient of 
the dark matter potential, i.e., the dimensionless function on the r.h.s. of 
equation (\ref{SVT}) {\it depends only on the galaxy structure} (see, e.g., 
Ciotti, Lanzoni, and Renzini 1996, CLR). Moreover, $\tilde\uss$ and 
$\tilde\wsh$ are known to be {\it weakly dependent} on the particular density 
profiles (see, e.g., Spitzer 1969, Ciotti 1991, Dehnen 1993). Obviously the 
same comments apply also to the stellar velocity virial velocity dispersion 
$<V_*^2>$, that necessarily results to be {\it independent} of the particular 
internal dynamics of the galaxy (e.g., the amount of orbital anisotropy). 

$<V_*^2>$ is related to $\ss^2$ through a dimensionless function that depends 
on the galaxy structure, its specific internal dynamics and on projection 
effects:
\begin{equation} 
<V_*^2>= \ck (structure, anisotropy,projection)\times\ss^2\enspace .
\label{fck}
\end{equation}
It is important to note that -- also in absence of orbital anisotropy -- $\ck$
{\it is very sensitive to galaxy-to-galaxy structural differences}, much more 
than $\tilde\uss$ and $\tilde\wsh$, because it relates a weakly structure 
dependent quantity ($<V_*^2>$) to a {\it local} one ($\ss^2$). So, in practice
structural non-homology always implies a {\it strong} variation in the 
$\sigma^2$ profile. Some authors call this phenomenon {\it dynamical} 
non-homology, but I (strongly!) suggest to call this effect {\it kinematical} 
non-homology, at least for two reasons. First, it is not very useful to call 
with two different terms (structural vs. dynamical non-homology) the 
{\it same} phenomenon: you cannot break structural homology without breaking 
also kinematical homology. Second, {\it dynamical} homology should be used to 
describe galaxies (for example) with the same relative amount of ordered 
rotation or orbital anisotropy. So, according to this nomenclature, 
kinematical non-homology can be induced by structural non-homology and/or 
dynamical non-homology. Moreover, globally isotropic galaxies with different 
density profiles are structurally non-homologous but dynamically homologous, 
their $\re<V_*^2>/G\ml\lb$ are similar, but their $\ck$'s can be significantly 
different. As a final remark, it is important to note that the strong 
dependence of $\ck$ on galaxy structure is essentially due to its definition: 
in fact, using larger and larger aperture velocity dispersions instead of the 
{\it central} velocity dispersion $\ss^2$, the projected Virial Theorem is 
better and better approximated, and for a spherical system without dark matter
$\ck\to 3$ independently of the galaxy orbital structure (e.g., Ciotti 1994; 
cfr. also the results of the numerical simulations of mergers of Capelato et 
al. 1995). 

From equations (2)-(3) and (5)-(6), defining 
\begin{equation}
\err\equiv [2\pi G\, (|\tilde\uss| + \mdr |\tilde\wsh|)]/\ck\enspace ,
\label{fer}
\end{equation}
one finally obtains 
\begin{equation}
\ku=\log(\err\times \ml\times\lb/2\pi)/\sqrt{2}\enspace ;\quad 
\kt=\log(\err\times \ml)/\sqrt{3}\enspace .\label{kteo}
\end{equation}

{\it Note that the VT does not imply any FP}, in fact for fixed $\lb$ 
different galaxies, all satisfying the VT, can in principle have very 
different $\err$ and $\ml$, and so be scattered everywhere in the $k$-space. 
So, the statement that {\it the FP deviates from the VT} -- often stated 
because of the difference between the exponents in eq. (\ref{VFP}) and those in
the VT [eq. (\ref{SVT})] -- is wrong: the FP deviates from {\it homology} (in
a broad sense). 

Summarizing, three ingredients are necessary for a class of hot dynamical 
systems to flatten about a FP: 1) to be virialized, 2) to have similar 
structures and internal dynamics, 3) to exhibit a small dispersion of $\ml$ 
for any given $\lb$. Observations tell us that $\err\times\ml$ is a very well 
defined function of the galaxy parameters with an intrinsic nearly constant 
scatter less than $12\%$. In particular, 
\begin{equation}
\delta (\err\times\ml)/(\err\times\ml)< 0.12\enspace .
\end{equation}
As galaxies in the BBF sample span a factor $\sim 200$ in $\lb$, the tilt 
corresponds to a factor $\sim 3$ increase of $\err\times\ml$ along the FP, from
faint to bright galaxies. If $\ml$ and $\err$ are not finely anticorrelated, 
this implies a very small dispersion, {\it separately} for both quantities, at
any location on the FP. This sets a very severe restriction on $\err\times\ml$,
which translates into strong constraints on the range that each parameter can 
span at any location on the FP. It is evident that fine tuning is required to 
produce the tilt, and yet preserve the tightness of the FP. 
Note also how, from eqs. (\ref{kteo}), galaxies with fixed $\lb$ and 
various internal dynamics, structure, $\ml$, {\it move along straight lines}
in the $k$-space, with $\kt=\sqrt{2/3}\ku+{\rm const.}$. The inclination of 
this line with respect to the FP given by BBF is equal to $\arctan(\sqrt{2/3})-
\arctan(0.15)\simeq 30\deg$: this is the reason why in numerical 
simulations the end-products of the merging of systems initially placed on the
FP are found near the FP itself\footnote{Moreover, this relation between $\kt$ 
and $\ku$ helps reduce slightly the problem of the FP thickness.}.
In conclusion, what is important is not the attempt to understand, perhaps by 
the finding of a "good" set of observational quantities, why the FP is 
"distant" from the VT, but, on the contrary, why galaxies are so similar in 
structure and dynamics, with such a small scatter.

\section{Exploring Various Possibilities}

For simplicity the origin of the FP tilt can be sought in two {\it orthogonal}
directions: either due to a {\it stellar population} effect, in which case 
$\ml\propto \lb^{0.2}$ and $\err$=const, or due to {\it structural/dynamical} 
effects, i.e., $\err\propto \lb^{0.2}$ and $\ml$=const. 

\subsection{A stellar origin: changing the IMF}

A systematic change of the stellar initial mass function (IMF) is explored 
in Renzini and Ciotti (1993, hereafter RC). $\ml$ is obtained by convolving 
the present mass of the stars $M$ with the IMF, where $M=\mi$ for the initial 
mass $\mi<\mto$ (the turnoff mass), and $M=\mr$ (the remnant mass) for 
$\mi\ge\mto$. For the IMF we adopt $\psi(\mi)=A\lb\mi^{-(1+x)}$, where $\lb$ 
is the present day blue luminosity of the population, or a multi-slope Scalo 
IMF with a variable slope for $M<0.3\msun$. For details see RC.

\subsubsection{Changing $\min$.}

In this case we assume a {\it decrease} of $\min$, the lowest stellar mass, 
for increasing $\lb$. We found that -- in the case of a single-slope IMF -- 
for no value of $x$ small values of $\ml$ (characteristic of the FP faint-end)
are obtained, unless $\min$ is unrealistically high. Reducing the slope does 
not help: for $x$ below $\sim 0.65$ $\ml$ increases again, since then the 
mass in remnants increases more than how much the mass in the lower main 
sequence stars is reduced. Only with the multi-slope Scalo IMF a low $\ml$ can
be realized (Fig. 1 in RC).

\subsubsection{Changing the IMF slope.}

The previous requirement of a low $\ml$ at the FP faint-end forces the choice 
of a low $x$, but then $\ml$ is quite insensitive to variations of $\min$; 
only for a steep IMF $\ml$ is sensitive to $\min$. As a consequence a mere 
variation of $\min$ with a constant IMF slope cannot account for the observed 
trend. Hence, a variation of slope is required, by an amount $\Delta x$ which 
depends on the adopted $\min$. We conclude that {\it a major change of the IMF
slope in the lower main sequence is necessary to account for the FP tilt}. 
There remains to consider the thickness of the FP. In RC it is shown that in 
order to preserve the $\sim 12\%$ upper limit on $\sigma(\ml)$, the 
galaxy-to-galaxy dispersion in $\min$ and $x$ should be extremely small, 
$<\pm 10\%$ and $<\pm 0.15$, respectively. Such very small galaxy-to-galaxy 
dispersion, coupled to a large systematic variation of $x$, is a rather 
demanding constraint, and we conclude that fine tuning is required to produce 
the observed tilt of the FP, while preserving its constant thickness: the IMF 
should be virtually universal for a given galaxy mass, and yet exhibit a large
trend with galaxy mass. A more accurate analysis of this scenario, using 
stellar population synthesis models, is given by Maraston (1996).

\subsection{A structural/dynamical origin}

In this case, assuming $\ml$ =const., we explore under which conditions 
structural/dynamical effects may cause the tilt in $\kt$ via a systematic 
increase of $\err$ [RC; CLR; Ciotti and Lanzoni 1996, (CL); Lanzoni and Ciotti
1996 (LC)]. In all these investigations spherical, non rotating, 
two--component galaxy models are constructed, where the light profiles 
resemble the $R^{1/4}$ law when projected. The internal dynamics is varied 
using the Osipkov-Merritt formula. Here for shortness reasons only the main 
results are summarized.

\subsubsection{Dark matter content and distribution.}

In this case we assume {\it global isotropy}, i.e., all models are 
{\it dynamically homologous}, and we ascribe all the FP tilt to systematic
variations of $\mdr=\mh/\ms$ or $\beta=\rch/\rcs$ ($\rch$ is a characteristic
radius of the dark matter distribution). 
Obviously, the larger $\beta$, the larger the 
variations of $\mdr$ that are required to produce the tilt. For 
Hernquist+Hernquist models and Hernquist+Plummer models, for $\beta \simeq 5$ 
exceedingly large values of $\mdr$ are required to produce the FP tilt 
($\mdr\simeq 30-175$). An increasing $\mdr$ may be at the origin of the 
observed tilt, provided that $\beta <5$. The same problem affects all the 
Jaffe+Quasi-isothermal models that we have considered, for every values of 
$\beta$ and $\rt$. For Jaffe+Jaffe models, $\mdr$ never becomes larger than 10
thus every value of this parameter is acceptable, for every explored value of 
$\beta$.

\subsubsection{Structural non-homology.}

Systematic deviations of the Es light distribution from the standard $R^{1/4}$
profile may also possibly cause the FP tilt (Djorgovski 1995; Hjorth and 
Madsen 1995). We investigate this {\it morphological} option using isotropic 
$\dvm$ models without dark matter, thus ascribing to a systematic variation 
of $m$ the 
origin of the tilt. Assuming the faintest galaxies to be $R^{1/4}$ systems, in 
order to produce the tilt $m$ has to increase from 4 up to $\sim 10$ along 
$\ku$ (CLR). If one assumes $m=2$ for the faintest galaxies, the required 
variation is even larger, about a factor of 4, up to $\sim 8$, and its 
permitted variation at each FP location remains very small: a scatter of $m$
$<10\%$ at any location on the FP should be associated to a large variation of
it with galaxy luminosities. A systematic trend of $m$ with galaxy luminosity 
has been reported (Caon, Capaccioli and D'Onofrio 1993), with $m$ increasing 
from $\sim 1$ up to $\sim 15$, thus spanning a much wider range than required 
to produce the tilt. We conclude that further observational studies are 
required in order to determine whether a progression of light-profile shapes 
along the FP really exists among cluster ellipticals.

\subsubsection{The role of anisotropy.}

In this case we ascribed the entire tilt of the FP to a trend with $\lb$ in 
the anisotropy degree of the galaxies (described in the Osipkov-Merritt 
formulation by the anisotropy radius $\ra$), assuming no dark matter. For 
Hernquist, Jaffe, and $\dvm$ models constrained to the FP we found that, above
a certain luminosity, the phase--space distribution function runs into 
negative values. So we conclude that anisotropy alone cannot be at the origin 
of the tilt, because the extreme values of $\ra$ that would be required 
correspond to dynamically inconsistent models (CLR, CL, LC). A special problem
with anisotropy is raised by the FP thinness: for galaxies with a positive 
distribution function a strong fine tuning between $\lb$ and $\ra$ could appear
to be required. However in CL and LC we showed, using a semi--quantitative 
global instability indicator, that in $\dvm$ models the excursion in 
anisotropy permitted by {\it stability} is much less than that given by the
simple dynamical consistency, and the induced scatter on the FP is inside the 
observed spread in $\kt$. We are well aware that this result is very 
qualitative (where is it placed in the $k$--space the end-product of a radially
unstable galaxy model initially on the FP?), and so we feel that this result 
is worth to be further studied, using N-body simulations.

\section{Conclusions}

Our exploration indicates that all structural/dynamical/stellar population
solutions to the FP tilt are rather unappealing, though some are more so than
others. This comes from the strong {\it fine tuning} that is required, no 
matter whether the driving parameter is the amount of dark matter ($\mdr$), 
its distribution relative to the bright matter ($\beta$), the shape of the 
surface brightness distribution ($m$), or finally the properties of the IMF. 
In addition to this, we have excluded a trend in the anisotropy as possible 
cause of the tilt, because it leads to physically inconsistent models. Finally,
we showed in a semi-quantitative way that probably a fine tuning of anisotropy
with the galaxy luminosity is not required for stability arguments, but 
deeper investigations, both analytical and numerical, are required.  
%
%

\end{document}


\pagenumbering{arabic}
\title{The Fundamental Plane of Galaxy Clusters}

\author{Alberto\, Cappi\inst{1} and
Sophie\, Maurogordato\inst{2}}

\institute{Osservatorio Astronomico di Bologna, via Zamboni 33,
I-40126 Bologna, Italy
\and
CNRS, LAEC, Observatoire de Paris-Meudon, 5 Place J.Janssen,
F-92125 Meudon Cedex, France}

\maketitle

\begin{abstract}{\baselineskip 0.4cm 
In the three--dimensional space defined by the logarithms of
central velocity dispersion $\sigma$, effective radius $R_e$ and
mean effective surface brightness $I_e$, elliptical galaxies 
are confined in a narrow plane (Dressler et al. 1987; Djorgovski 
\& Davis 1987). Here we discuss the observational evidence
for the existence of an analogous relation for galaxy clusters
(Schaeffer et al., 1993).
}
\end{abstract}

The relations between global observables in stellar systems are important
both from a theoretical and a practical point of view.
The Tully--Fisher relation for spirals
and the Faber--Jackson relation for ellipticals involve two observables.
In the case of ellipticals, the residual scatter suggested 
the introduction of a third parameter, resulting in the definition of 
the so--called fundamental plane (Djorgovski \& Davis 1987; 
Dressler et al. 1987).
From the virial theorem we expect a relation involving the three observables
$R$, $L$ and $\sigma$, which can be expressed in the following way 
(see Djorgovski \& Santiago 1993):
$ L \propto K R \sigma^2 (M/L) ^{-1} $
where $K$ is a structural parameter. 
Therefore the existence of the FP for a given class of objects
is not a trivial consequence of the virial
theorem; it requires also that the class of objects under study has
a similar dynamical structure and a tight mass to light ratio 
with a small dispersion.

We decided to search for relations between global observables in galaxy 
clusters (Schaeffer et al. 1993).
We used the effective radii and total luminosities for a sample of 29 
{\em regular} galaxy clusters, measured after an accurate and homogeneous 
reduction of high--quality photometric data by West et al. (1989; WOD), who
had found a well--defined radius--luminosity relation, $R \propto L^{0.5}$.
For 16 of these clusters we found reliable measures of velocity
dispersion (Struble \& Rood 1991).
Combining these data we showed the existence of a relation 
between velocity dispersion and luminosity, $L \propto \sigma^{1.9}$.
The relation $L$--$\sigma$ for clusters is the equivalent of the
Faber--Jackson law for ellipticals, but in clusters
the luminous matter in stars is a
small fraction of the total mass, 
dominated by dark matter and the hot gas.
We realized also that
the relations $R$--$L$ and $L$--$\sigma$ had a scatter
larger than one could expect from observational errors.
Introducing $\sigma$ as a third parameter, we
found that these clusters define a Fundamental Plane
(fig.\ref{fig:fig1}), as many stellar systems from globular clusters 
(Nieto et al. 1990) to elliptical galaxies 
(see fig.3 in Schaeffer et al. 1993).
We found $ L \propto R^{0.89 \pm 0.15} \sigma^{1.28 \pm 0.10}$ or, defining
a surface brightness $I_e = L/R_e ^2$,
$R_e \propto {I_e}^{-0.81} \sigma ^{1.15}$,
with the best fit parameters quite similar to 
the elliptical ones. The cluster FP is obviously shifted relatively to
the elliptical one, because of the different $M/L$ ratio.
Therefore all gravitationally bound systems appear to define their FP.
The common link is the virial theorem, but of course the slope
and dispersion of each plane will depend on the class of objects taken
into account. Might the cluster FP be the result of spurious effects?
Discussing their $R$--$L$ relation for galaxy clusters,
WOD excluded a bias
due to a selection in surface brightness, as one Abell radius
includes most of the cluster luminosity. 
We also note that cluster peculiar velocities
cannot produce the observed relation. Furthermore,
tests with numerical simulations (Pentericci et al. 1996)
show that the cluster
FP is conserved after merging and that it cannot be spuriously generated
by the procedure used to find $R_e$ and $L$.

One can also use the coordinates defined
by Bender et al. (1992) (fig.\ref{fig:fig1}).
These coordinate system, which does not correspond exactly to the FP, 
gives $k_3 \propto \log(M/L)$ vs. $k_1 \propto \log(M)$. 
It is clear that the $M/L$ of clusters has a small dispersion
(see the discussion in Renzini \& Ciotti 1993).
We find a small trend of $M/L$ with $L$, 
with $M/L \propto L^{0.3 \pm 0.1}$.

We can also have a rough estimate of the cluster peculiar velocity 
$V_{p} = V_{obs} - H_0 D$ (assuming that the deviation from the
FP is entirely due to $V_p$, and neglecting the scatter of the $M/L$ ratio).
Uncertainties become very large beyond $z=0.05$; for
the 10 clusters at $z \le 0.05$ we find (Cappi et al. 1994)
$ V_p \le 1000$~km/s, with a trend as a function
of the cosine angle between clusters
and the GA direction consistent with the results of Han \& Mould (1992);
the upper limit $\Delta H / H \le$ 15\% is consistent with 
Lauer \& Postman (1992).

We conclude that regular clusters define
a FP comparable to that of elliptical galaxies,
which can give us information
about formation time dispersions, $M/L$ ratios, or cluster $v_p$.
A critical point is the quality of the data,
especially the photometry. 
For example, a number of recent data give measures of virial radii 
significantly underestimated, because based on a limited 
region of the cluster (as discussed by Carlberg et al. 1996). 
One should remind that typical effective radii are around 1 \h, 
while virial radii are more than 2 times $R_e$.
Therefore a large observational effort is still required to address many 
important issues, and we need CCD photometric surveys covering a large part 
of each cluster and not only its central regions.
Last but not least, the combination with X--ray data will be extremely
useful to understand the role of the different matter components 
in galaxy clusters.

%
%
\begin{figure*}[ht]
\epsfxsize=12cm
\epsfbox{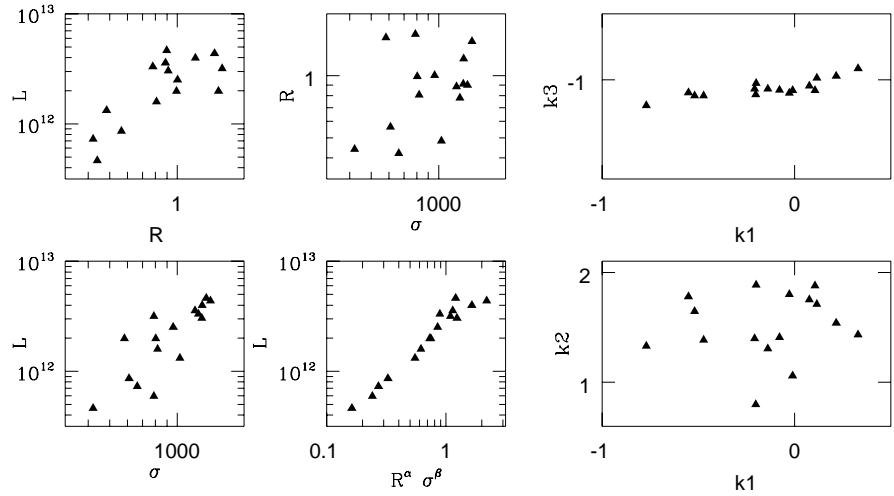}
\caption[]{Relations between $L$, $R$ and $\sigma$, and between $k_1$,
$k_2$ and $k_3$}
\label{fig:fig1}
\end{figure*}


\pagenumbering{arabic}

\def\r1m{$R^{1/m}$}
\def\dvm {R^{1/m}}
\def\Iz {I_{\circ}}
\def\re {R_{\rm e}}
\def\ra {r_{\rm a}}
\def\sac{(\ra)_{\rm c}}
\def\rax{(\ra)_{\xi}}
\def\sp {\sigma_{\rm P}}
\def\srad{\sigma_{\rm r}}
\def\stan{\sigma_{\rm t}}
\def\sap{\sigma_{\rm a}}
\def\vrad{v_{\rm r}}
\def\vtan{v_{\rm t}}
\def\svir{\sigma_{\rm V}}
\def\Krad{K_{\rm r}}
\def\Ktan{K_{\rm t}}
\def\e {{\cal E}} 
\def\Qt{\widetilde Q}
\def\hq{h_4}
\def\hs{h_6}
\def\ltot {L_m}
\def\ltota {\widetilde\ltot}
\def\pot {\psi _m}
\def\pota {\widetilde\pot}
\def\ml {\Upsilon}
\def\dm {\widetilde\nu _m}
\def\dfm{f_m}
\def\dfmt{\widetilde\dfm}
\def\dfqt{\widetilde {f_4}}
\catcode`\@=11
\def\gsim{\ifmmode{\mathrel{\mathpalette\@versim>}}
    \else{$\mathrel{\mathpalette\@versim>}$}\fi}
\def\@versim#1#2{\lower 2.9truept \vbox{\baselineskip 0pt \lineskip
    0.5truept \ialign{$\m@th#1\hfil##\hfil$\crcr#2\crcr\sim\crcr}}}
\catcode`\@=12
\title{Anisotropic $R^{1/m}$ Models. Velocity Profiles, and the FP of 
Elliptical Galaxies} 

\author{Barbara Lanzoni\inst{1,\ast} and Luca Ciotti\inst{2}}

\institute{Dipartimento di Astronomia di Bologna, Via Zamboni 33, Bologna, Italy
\and
Osservatorio Astronomico di Bologna, Via Zamboni 33, Bologna, Italy}

\maketitle

\begin{abstract}
We study the dynamical properties of spherical galaxies with surface 
luminosity profile described by the $\dvm$-law, in which a variable degree 
of orbital anisotropy is allowed. The stability of the models against 
radial--orbit instability is studied, and the limits on the maximum anisotropy
allowed for each model are determined. The consequent constraints on the the 
projected velocity dispersion imply that no fine-tuning for anisotropy is 
required along the fundamental plane (FP) in order to maintain its small 
thickness. Finally, the velocity profiles are constructed, and their 
deviations from a gaussian discussed.
\end{abstract}

\section{The models}
The models surface brightness profile is described by the $\dvm$ law 
(Sersic 1968): 
\begin{equation}
I(R)=\Iz \exp{\left[-b(m)\left(R/\re\right)^{1/m}\right]}, \label{r1m}
\end{equation}
which seems to give a better fit to the spheroidal galaxies surface brightness 
profiles (e.g., Caon et al. 1993; Graham et al. 1996, and references therein)
than the ``standard" $R^{1/4}$ law [de Vaucouleurs 1948; Eq.(1) for $m=4$].

The velocity dispersion tensor of our models is described by the 
Osipkov--Merritt parameterization (Osipkov 1979; Merritt 1985), and 
characterized by the {\it anisotropy radius} $\ra$: 
in the limit $\ra\to\infty $ the velocity dispersion tensor is globally 
isotropic, while in general the radial anisotropy increases with radius. 

\section{Stability}
The stability of the anisotropic models is investigated in a semi-quantitative
way using the radial-orbit instability indicator $\xi$ (e.g., Fridman and 
Polyachenko 1984). A model is likely to be unstable if 
\begin{equation}
\xi\equiv {2\Krad\over\Ktan} \gsim 1.5\div 2,\label{stab}
\end{equation}
where $\Krad$ and $\Ktan$ are the radial and the tangential kinetic energies, 
respectively. 
This parameter is quite independent of the assumed density distribution 
profile, and for any globally isotropic system $\xi=1$, 
while in presence of radial anisotropy $2\Krad >\Ktan$, and so $\xi>1$. 

\paragraph{$^\ast$}Present address: 
Osservatorio Astronomico di Capodimonte, Via Moiariello 16, Napoli, Italy.
\vfill\eject

For the investigated models, $\xi=\xi(\ra,m)$, and for a fixed $m$ it decreases
towards unity for increasing $\ra$ (see {\bf Fig.1}), according to 
the previous discussion. So, assuming a fiducial critical value of $\xi$ for
stability (e.g., $\xi=1.7$), a minimum value for the anisotropy radius 
$\rax$ is obtained, i.e., all models with $\ra <\rax$ are unstable. 

\begin{figure}
\hskip 2cm
\psfig{figure=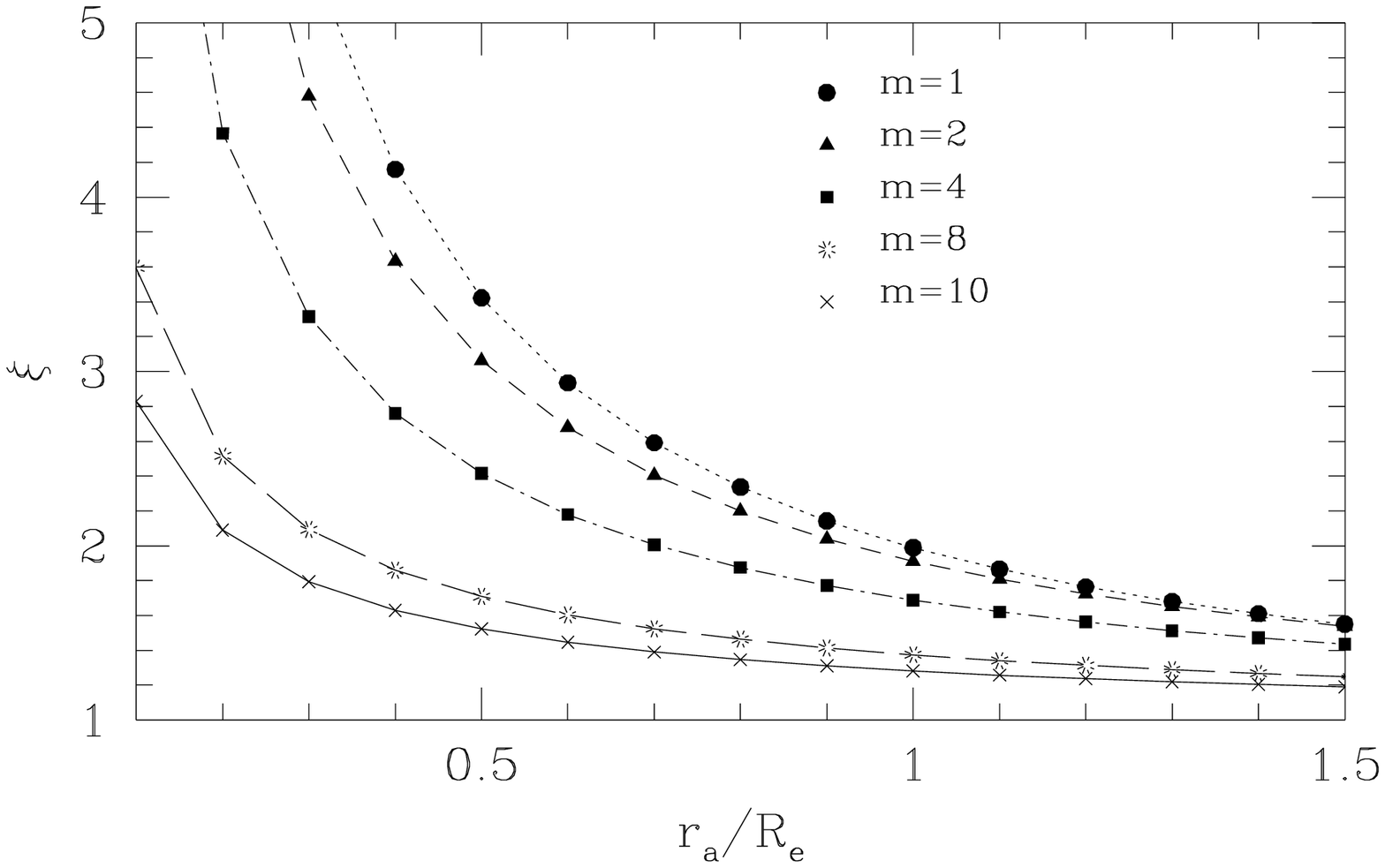,width=6cm,height=4cm}
\end{figure}

For stable models we compute the spatial velocity dispersion profile by 
integration of the Jeans equation. After the operations of 
projection and mean, we obtain the {\it aperture} velocity 
dispersion $\sap$, which mimics the observed central one (see Ciotti and 
Lanzoni 1997).

\section{Implications for the FP}
In principle, one of the possible origins of the FP tilt could be a 
systematic increase of radial anisotropy from faint to bright galaxies 
(Ciotti et al. 1996).
If this is the case, a variation of factor 3 from isotropic to 
anisotropic squared central velocity dispersion $\sap^2$ is required, 
in the assumption of structural homology (i.e., the same $m$ for all galaxies). 
On the contrary, for our models we find that, if the maximum degree of 
anisotropy allowed by the stability requirement is considered, the 
{\it radial anisotropy cannot produce the tilt}   
(a conclusion already reached for different galaxy models in Ciotti et al. 
1996).

As concerns the problem of the very small thickness of the FP, 
the limits imposed by the stability requirement imply that 
the variations between the isotropic and the maximum anisotropic velocity 
dispersion are so small that the {\it anisotropy is not 
required to be fine-tuned with the galaxy luminosity in order to 
maintain the small observed FP scatter} (see Ciotti and Lanzoni 1997).

\section{Velocity Profiles}
The velocity profile (VP) at a certain projected distance from 
the galaxy center is the distribution of the stars line--of--sight velocities 
at that point. The analysis of the deviations of VPs from gaussianity may give
important insights on the dynamical structure of a galaxy (e.g., van der 
Marel 1994).

We numerically recover the VPs of our models (as in Carollo et al. 1995), and  
expand them on the Gauss-Hermite basis (Gerhard 1993; van der Marel and 
Franx 1993), thus obtaining 
the values of the coefficient $\hq$ at various distances from the galaxy 
center. 
Generally, a negative $\hq$ indicates a flat--topped VP, while a 
positive one indicates a VP more centrally peaked than a Gaussian. 

In {\bf Fig.2} the radial behaviour of $\hq$ for various $m$ and for 
isotropic (left panels) and anisotropic ($\xi=1.7$, right panels) models is 
shown.

\begin{figure}
\hskip 2cm
\psfig{figure=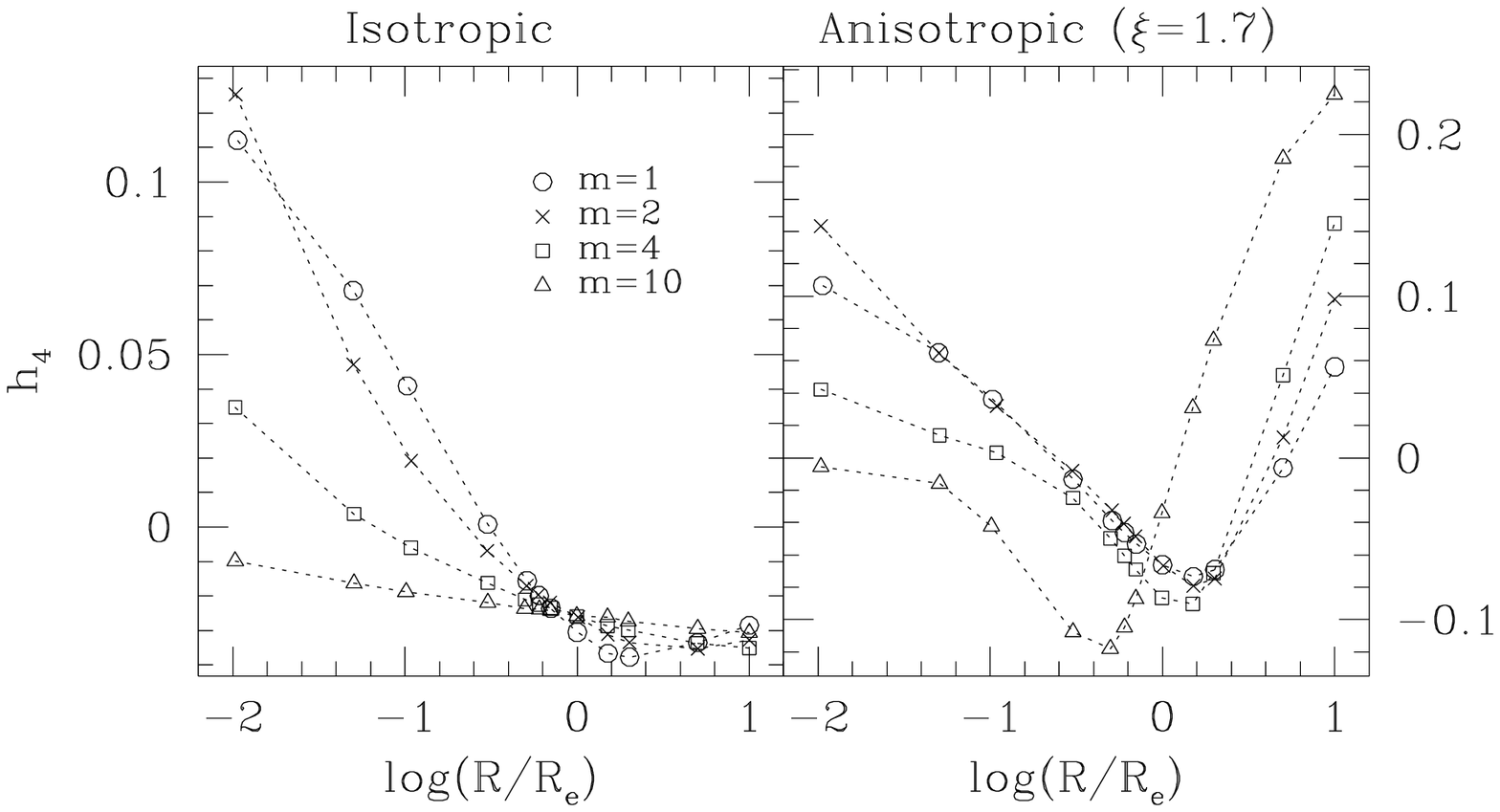,width=7cm,height=4cm}
\end{figure}

Because of the differences in the trend of $\hq$ between the isotropic and the
anisotropic cases, and because of the growing evidence that $\dvm$ law 
appropriately describe the surface brightness profiles of elliptical galaxies,
we suggest that a detailed study of the VPs along the FP could be in 
principle a tool to study the effect of orbital anisotropy on its tilt and 
thickness.


\pagenumbering{arabic}
\title{The Relation between X-ray Emission,\\
Galaxy Structure and Internal Kinematics \\
in Early-type Galaxies}

\author{Silvia\,Pellegrini}

\institute{Dipartimento di Astronomia, Universit\`a di Bologna,\\
via Zamboni 33, I-40126 Bologna, Italy }

\titlerunning{X-ray Emission, Galaxy Structure and Internal Kinematics }
\maketitle

\begin{abstract}
Using data in the literature plus new spectroscopic observations, we have built 
the first large sample of X-ray emitting early-type galaxies with 
known kinematics (central velocity dispersion $\sigma_{\rm c} $ and maximum 
rotational velocity $v_{\rm rot}$). With this sample we
investigate the effect of rotation on the X-ray emission, particularly with 
regard to the X-ray underluminosity of flat systems. 
The role of structural parameters as the ellipticity $\epsilon$, the isophotal
shape parameter $a_4/a$, and the inner slope of the surface brightness profile
$\gamma$ recently measured by $HST$ are also investigated. 
\end{abstract}
\section{The Problem}
The large scatter in the $L_{\rm X}-L_{\rm B}$ diagram is the most striking 
feature of the X-ray properties of early-type galaxies: it well reaches two 
orders of magnitude in $L_{\rm X}$ at any fixed $L_{\rm B} > 3\, \, 10^{10}
L_{\sun} $ (Fabbiano et al. 1992).  It was originally explained by 
environmental differences (White \& Sarazin 1991), or by different dynamical 
phases for the hot gas flows, ranging from winds to subsonic outflows to 
inflows (Ciotti et al. 1991). Recent observational results have produced a new
debate on the explanation of the scatter. Eskridge et al. (1995a,b) showed that
on average S0s have lower $L_{\rm X}$ and $L_{\rm X}/L_{\rm B}$ at any fixed 
$L_{\rm B}$ than do Es. Moreover, galaxies with axial ratio close to unity span 
the full range of $L_{\rm X}$, while flat systems all have $L_{\rm X}\la 10^
{41}$ erg s$ ^{-1}$; this holds both for Es and S0s separately.
Has the shape of the mass distribution any role in determining $L_{\rm X}$?
Since flat galaxies exhibit a higher rotation level, what is the effect of 
galactic rotation?

Finally, recent $HST$ results show that
hot galaxies can be divided into two types: 
{\it core} galaxies, described by a surface brightness profile $I\propto r^{-
\gamma}$ ($\gamma\leq 0.3$) at the center, and {\it power-law} galaxies, with 
steep, featureless profiles (Faber et al. 1996). 
These central properties correlate with global ones as rotation and isophotal 
shape. Is there any correlation also with global $L_{\rm X}$?

\section{Our observations and sample }
We obtained measurements of velocity dispersion $\sigma$ and radial velocity 
$v$ along the major axis for 7 Es and S0s belonging to the {\it Einstein} 
sample (Fabbiano et al. 1992), the largest homogeneous sample of early-type 
galaxies with measured X-ray emission. The main characteristics of these 
galaxies, the details of the spectroscopic observations, the final radial 
velocity and velocity dispersion profiles, and the kinematic properties of the
final sample obtained by collecting data also from the literature 
are given in Pellegrini et al. (1996, hereafter PHC). 

\section{Results}
\subsection{X-ray emission, rotation and flattening}
The trend between the X-ray to optical ratio $L_{\rm X}/L_{\rm B}$, a measure 
of the hot gas content of the galaxies, and $v_{\rm rot}/\sigma_{\rm c}$ is 
L-shaped, with the X-ray brightest objects confined at $v_{\rm rot}/\sigma_
{\rm c}\la 0.4$, both for Es and S0s (Fig.~1, with $L_{\rm X}$ and $L_{\rm 
B}$ derived as in PHC). The trend between $L_{\rm X}/
L_{\rm B}$ and the ellipticity $\epsilon$ is also L-shaped, even though less 
sharp:
there are no high $L_{\rm X}/L_{\rm B}$ objects with high $\epsilon$. Instead,
there is no clear trend between $L_{\rm X}/L_{\rm B}$ and the 
anisotropy parameter $(v/\sigma)^*$.
So, the observations suggest that the major effect on $L_{\rm X}/L_{\rm B}$ is 
given by the relative importance of rotation and random motions (see also PHC).

\begin{figure*}
\epsfxsize=12.1cm
\hoffset=-1cm
\epsfbox{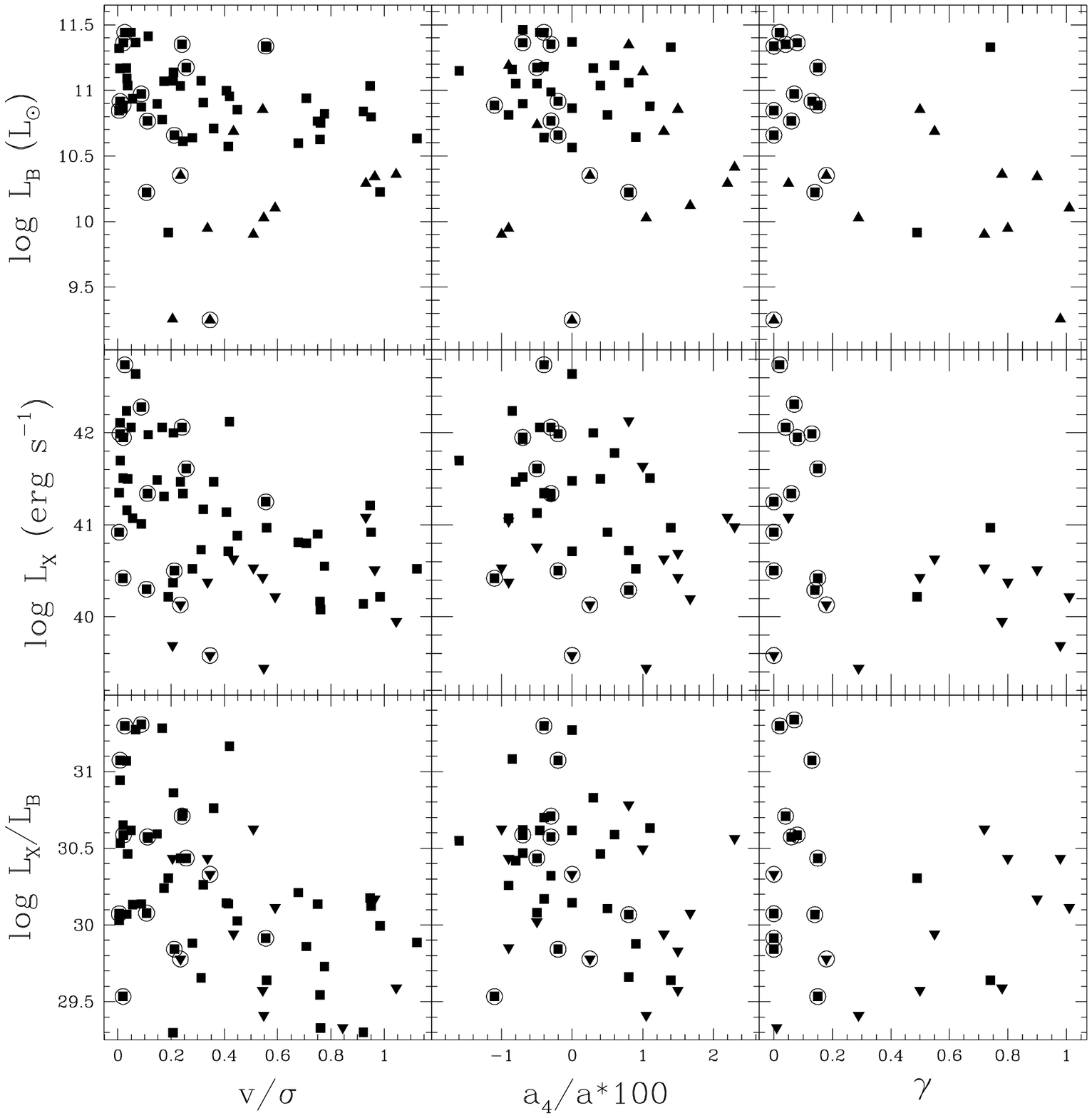}
\caption[]{
$a_4$ is from Eskridge et al. (1995b), $\gamma$ from Faber et al. (1996). 
Boxies are X-ray detections, triangles
upper limits;  {\it core} galaxies have been circled.   }
\end{figure*}

\subsection{X-ray emission, isophotal shape, and central 
brightness profile}
Faber et al. (1996) show that 
core galaxies tend to be boxy and slowly rotating, whereas power-law
galaxies tend to be disky and rapidly rotating.
Bender et al. (1989) showed that disky objects are low X-ray emitters, 
while boxy and irregular objects show a large range of $L_{\rm X}$.

The results of our investigation (Fig.~1) show a striking
relationship between $L_{\rm X}$ and $\gamma$:
core galaxies span the whole range of $L_{\rm 
X}$, while power-law galaxies are confined within  log$L_{\rm X}\la 41$.
The origin of this new threshold effect should be looked for in the connection
between central structural properties (a central black hole? different
formation histories?) and global $L_{\rm X}$.

\section{Conclusions}
Rotation and flattening are predicted to have different effects in the pure 
inflow and in the wind/outflow/inflow scenarios. Rotation plays the major role
 in the first one (Brighenti \& Mathews 1996), but 
the expected correlation of $L_{\rm X}/L_{\rm B}$, or 
of the residuals 
of the $L_{\rm X}-L_{\rm B}$ correlation, with $v_{\rm rot}/\sigma_{\rm c}$ 
is not present (PHC). Flattening  is 
determinant in the second scenario (Ciotti \& Pellegrini 1996; D'Ercole \& 
Ciotti 1996), but it remains to be explained 
why the L-shape is more pronounced with respect to $v_{\rm rot}/\sigma_{\rm c}$
 rather than to $\epsilon$. 

The most significant effect found is the L-shape 
in the $L_{\rm X}-\gamma$ plot: the division of early-type galaxies into two 
types
corresponds to a difference also in their X-ray properties. This represents 
another link between the central
structures of hot galaxies and their global properties, that might also be
useful to better understand their formation history and evolution.

\voffset=-0.5cm